\documentclass[aip,amsmath,amssymb,reprint,author-year]{revtex4-1}
\usepackage{amsmath,amssymb}
\usepackage{graphicx}
\usepackage{dcolumn, xcolor}%
\allowdisplaybreaks
\usepackage{bm}

\newcommand{\dd}{\mathrm{d}}

\newcommand{\mnras}{Mon. Not. Roy. Astron. Soc.}

\providecommand\apss{Astrophys. Space Sci.}
\providecommand\apj{Astrophys. J. }
\providecommand\apjl{Astrophys. J. Lett.}

\begin{document}
\title[]{\textbf{Relativistic regularized kappa distributions}}
\author{Linh Han Thanh}
\email {lht@tp4.rub.de}
\affiliation{Institut f\"ur Theoretische Physik IV, Ruhr-Universit\"at Bochum, 
Universit\"atsstrasse 150, 44780 Bochum, Germany}
\author{Klaus Scherer}
\affiliation{Institut f\"ur Theoretische Physik IV, Ruhr-Universit\"at Bochum, 
Universit\"atsstrasse 150, 44780 Bochum, Germany}
\affiliation{Research Department Plasmas with Complex Interactions,  Ruhr-Universit\"at Bochum, 
Universit\"atsstrasse 150, 44780 Bochum, Germany}
\author{Horst Fichtner}
\affiliation{Institut f\"ur Theoretische Physik IV, Ruhr-Universit\"at Bochum, 
Universit\"atsstrasse 150, 44780 Bochum, Germany}
\affiliation{Research Department Plasmas with Complex Interactions,  Ruhr-Universit\"at Bochum, 
Universit\"atsstrasse 150, 44780 Bochum, Germany}

\begin{abstract}
The special relativistic generalization of isotropic regularized kappa distributions is derived and compared to that of the original Olbertian (or standard) kappa distributions. It is demonstrated that for the latter the kappa parameter is even stronger limited than in the non-relativistic case, while for the former all positive kappa values remain possible. After a derivation of the non-relativistic limits, the pressures of the distributions are studied as a specific case of the moments of both the relativistic standard and regularized kappa 
distributions.  
\end{abstract}
\maketitle

\section{Introduction and motivation}
Many space physical and astrophysical plasmas are not in thermal 
equilibrium, so that their velocity (or momentum) distribution functions
deviate significantly from a Maxwell-Boltzmann distribution (MBD). 
For the quantitative treatment of moderately non-thermal particle 
populations, often referred to as suprathermal particles, the family 
of so-called kappa distributions has been proven extremely useful 
\citep[e.g.,][]{Lazar-Fichtner-2021}. This is because, on the one hand, 
these distributions allow for a simultaneous modelling of the power law 
wings and of a Maxwellian core of observed distributions and, on the 
other hand, because they contain the important reference case of an 
equilibrium Maxwellian as an asymptotic limit. 

In recent years the particular significance of regularized kappa 
distributions (RKDs) has been demonstrated for various cases. First, after
having defined and validated the RKDs, \citet{Scherer-etal-2018} showed
that they remove all divergences occurring when using the original
Olbertian \citep{Olbert-1968} or a modified version 
\citep{Matsumoto-1972} - both are referred to as standard kappa 
distributions (SKDs) in the following - and that they extend the 
range of permitted kappa values to $\kappa > 0$. Second,  \citet{Fichtner-etal-2018} demonstrated that RKDs 
are consistent with an extensive entropy unlike other kappa 
distributions \citep[e.g.,][]{Silva-etal-1998}. Third,  \citet{Lazar-etal-2020} employed the RKDs for an improved treatment 
of heat conduction in the solar wind plasma. And, fourth, 
\citet{Husidic-etal-2020} studied various wave modes within a 
RKD-based linear dispersion theory. Following these 
theoretical considerations, their usefulness for the analysis of
spacecraft data has been demonstrated by \citet{Scherer-etal-2021}.
For a comprehensive overview on the theory and applications of RKDs
and related distributions, see \citet{Lazar-Fichtner-2021}. 

Despite these improvements of the theory of kappa distributions, 
it was also demonstrated that even for RKDs there can exist 
significant unphysical contributions to the velocity moments 
resulting from superluminal particles with speeds above the speed of 
light, which are formally present in any non-relativistic theory
\citep{Scherer-etal-2019a}. While, in contrast to the case of SKDs, such unphysical contribution can be made negligibly small for RKDs
in case of non-relativistic plasma populations, the actual relativistic case remains problematic. 

With the present paper we intend to close this gap in the theory by 
defining special relativistic RKDs. Limiting the study to isotropy in
velocity or momentum space, we demonstrate that even in a relativistic treatment, such regularized distributions are mandatory, i.e.\ 
that, perhaps contrary to expectation, the use of relativistic SKDs
\citep[as considered in, e.g.,][]{Mann-etal-2006, Xiao-2006, Davelaar-etal-2019,Fahr-Heyl-2020}
is even more limited than that of non-relativistic ones, as the 
kappa value is even more constrained. 

Such distribution functions are subjects of relativistic kinetic theory 
to which several monographs have been devoted \citep[e.g.,][]{stewart, 
DeGroot, hakim} that summarize the state of the theory until the 
respective year of publication. When constructing the special relativistic 
generalizations of the non-relativistic counterparts, one must observe 
the proper limits. Besides the non-relativistic distributions, one 
important limit is the relativistic form of the MBD, i.e.\ the so-called
Maxwell-J\"uttner distribution (MJD), which represents the first step 
towards relativistic kinetic theory \citep{Juettner-1911}  and which 
has been used subsequently in various studies of plasma waves 
\citep[e.g.,][]{Fichtner-Schlickeiser-1995}, of the laser-induced acceleration of 
ions \citep{Huang-etal-2013}, of spontaneous electromagnetic fluctuations
\citep{Lopez-etal-2015} and other physical systems. In the present 
paper we demonstrate that the given definitions for relativistic SKDs and 
RKDs fulfil all relevant limits. 

Relativistic particle populations are found almost everywhere in the Universe. As a consequence of their low mass (as compared to ions),
particularly electrons often have to be described as relativistic populations, for example in the terrestrial 
radiation belts \citep{Grachi-etal-2021} and magnetosphere \citep{Kubota-etal-2015}, in solar flares \citep{Chen-etal-2020}, the solar wind \citep{Wang-etal-2015}, at the outer planets \citep{Roussos-etal-2016, Logachev-etal-2021}, or in the context of astrophysical turbulence \citep{Trotta-etal-2020}. For those populations the need of a relativistic 
modelling is evident. However, as outlined above, the latter is also 
required for the description of `merely' suprathermal particle populations. So, also a consistent description of suprathermal ion populations, which occur frequently, like pick-up ions in the solar wind \citep[e.g.,][]{Heerikhuisen-etal-2008, Zirnstein-McComas-2015, Fahr-etal-2016}, eventually needs to be done within a relativistic framework.  

Once the correct relativistic distribution functions have been defined, 
it is of interest to compute their velocity or momentum moments, which 
due to the relativistic cut-off at the speed of light exist for all orders, 
but for the SKDs, nonetheless, not for all kappa values. As for the distributions 
themselves, also these moments must have the correct limits. In particular, 
they must reproduce the moments of the MJD, which can be found, for example,  
in the textbook by \citet{hakim}. 

Before continuing with the construction of the desired relativistic 
distributions, we would like to emphasize that in the present first study 
of relativistic RKDs, we consider only isotropic distributions.  
Anisotropic versions as well as generalisations of RKDs have been studied in 
quite some detail in \citet{Scherer-etal-2019b, Scherer-etal-2021}, followed 
by applications to linear dispersion theory \citep{Husidic-etal-2020}, heat
conduction \citep{Lazar-etal-2020}, and data fitting \citep{Scherer-etal-2021}. Also, anisotropic generalizations of the MJD have been studied by various 
authors \citep[e.g.,][and references therein]{Yoon-1989, Schlickeiser-2004, Ahmad-Mahdavi-2016, Yoon-2007, Tautz-2010, Livadiotis-2016, Treumann-Baumjohann-2016}. Since, however, their structure is naturally more complex, the relativistic generalisations of anisotropic RKDs will be studied elsewhere. 

In the following we give a brief overview of the non-relativistic and 
relativistic MBD, SKDs and RKDs (section~II and section~III, respectively), study their asymptotic behaviour in section~IV,  
derive their moments in section~V, and briefly summarize and discuss all results in the final section~VI. 

\section{A brief overview of the non-relativistic distributions}
We briefly recapitulate the (isotropic) non-relativistic distribution functions of which we aim to study relativistic extensions. A plasma being in thermal equilibrium is described by the Maxwell-Boltzmann distribution (MBD)
 \begin{equation}
    f_{\textrm{MBD}}(v) = nN_{\textrm{MBD}}\exp{\left(-\frac{v^2}{\theta^2}\right)}, \label{eg:MBD}
 \end{equation}
 with $n$ denoting the number density, $N_{\textrm{MBD}}$ a normalization 
 constant, $v$ the particle speed, and $\theta=\sqrt{2k_BT/m}$ the most probable speed of the MBD. In the latter formula $k_B$ is the 
 Boltzmann constant, $T$ the temperature, and $m$ the particle mass. 
 
 The SKDs and RKDs are given by
  \begin{align}
    f_{\textrm{SKD}}(v)= &nN_{\textrm{SKD}}\left(1+\frac{v^2}{\theta^2\kappa}\right)^{-\kappa-1} \label{eq:SKD}\\
f_{\textrm{RKD}}(v)= &nN_{\textrm{RKD}}\left(1+\frac{v^2}{\theta^2\kappa}\right)^{-\kappa-1}\exp\left(-\alpha^2\frac{v^2}{\theta^2}\right) \label{eq:RKD}
 \end{align}
 describing an enhanced occurrence of suprathermal particles as 
 power laws extending from a Maxwellian core. The parameter $\kappa$ can be interpreted as a measure of the system's deviation from equilibrium in such a way that with $\kappa\rightarrow\infty$ one recovers the MBD. While the power law extends to infinity in case 
 of the SKDs, for the RKDs the additional parameter $\alpha\in [0,1]$ determines an exponential cut-off. For the special case $\alpha=0$ an RKD reduces to an SKD. While the above notation describing the dependence on speed is the most common one for the non-relativistic case, it is more appropriate to use the dependence on momentum for the relativistic case, which is being done hereafter. 
 
 With the normalization $n =4\pi\int f(p) p^2\dd p$, with $n$ being the particle number density, the normalization constants $N_{\ast}$ can be calculated to
 \begin{align}
    N_\textrm{MBD} = &\frac{1}{m^3\sqrt{\pi^{3}}\theta^3},  \\
    N_{\textrm{SKD}}=&\frac{1}{m^3\theta^3\sqrt{\pi^3\kappa^3}}\frac{\Gamma(\kappa+1)}{\Gamma(\kappa-\frac{1}{2})},\\
    N_{\textrm{RKD}} =&\frac{1}{m^3\theta^3\sqrt{\pi^3\kappa^3}U(\frac{3}{2},\frac{3}{2}-\kappa;\alpha^2\kappa)},
\end{align} 
where $U(a,b,z)$ %, or sometimes also denoted as $\Psi(a,b;z)$, 
is the Kummer-U function and $\Gamma(x)$ the Gamma function.
Note that, since the normalization is made in momentum space, a factor $m^{-3}$ appears in each $N_{\ast}$, which is absent when normalizing w.r.t.\ velocity. 

As recently pointed out by \citet{Scherer-etal-2019a}, there can be 
unphysical contributions to macroscopic quantities such as pressure and entropy due to particles with speeds $v$ greater than the speed of 
light $c$, which in the non-relativistic treatment are formally allowed as $v\in [0,\infty)$. 
In difference to an MBD, for which such contributions are generally 
negligible, for an SKD they become significant at the latest when the 
mean particle speed exceed $0.2\,c$. Although with an adjusted 
exponential cut-off this threshold can be increased for an RKD, 
for both kappa distribution types it becomes eventually necessary to 
switch to a relativistic description.  

\section{The relativistic distributions}
The relativistic generalization of the MJB, first derived by \citet{Juettner-1911}, reads
\begin{equation}
    f_{\textrm{MJD}}(p)=\frac{n\beta}{4\pi(mc)^3 K_2(\beta)}\exp(-\beta \gamma), \label{MJD}
\end{equation}
where $K_2(x)$ is the modified Bessel function of the second kind of order two, $\beta=mc^2/(k_BT)$, and the Lorentz factor $\gamma=\sqrt{1+p^2/(mc)^2}$. In the case of low temperatures, $\beta \gg 1$, and low speeds compared to that of light (a case 
formally describable by $c\rightarrow \infty$), the MJD reduces to 
the non-relativistic Maxwell-Boltzmann distribution.

A special relativistic version of SKDs has apparently
first been attempted by \citet{Naito-etal-1996}. Different versions have been suggested subsequently by, e.g., \citet{Xiao-etal-1998} and
\cite{Xiao-2006}, see \citet{Pierrard-Lazar-2010} for a brief overview.

The requirements for such distributions are as follows:
\begin{itemize}\setlength{\itemsep}{0.1cm}
    \item Lorentz invariance
    \item Recovery of the corresponding non-relativistic distribution functions in the limit of low temperatures ($\beta\gg 1$) and low speeds ($\gamma\approx 1$)
    \item Reduction to the equilibrium distribution, here the MJD, in the limit $\kappa\rightarrow \infty$
    \item Cut-off for speeds $v\rightarrow c$
\end{itemize}
Based on these requirements, we employ the following relativistic forms of the distributions:
\begin{align}
f_{\textrm{rSKD}}(p)&=nN_{\textrm{rSKD}}\left[1+\frac{\beta (\gamma-1)}{\kappa}\right]^{-\kappa-1}, \label{eq:rSKD}\\
f_{\textrm{rRKD}}(p)&=nN_{\textrm{rRKD}}\left[1+\frac{\beta (\gamma-1)}{\kappa}\right]^{-\kappa-1}
\!\!\!\!\!\!\!\!\!\!\exp\left[-\alpha^2\beta(\gamma-1)\right]
\end{align}
Note that this relativistic generalization of SKDs is consistent with that suggested by \citet{Xiao-2006}, \citet{Mann-etal-2006}, \citet{Davelaar-etal-2019}, and \citet{Fahr-Heyl-2020}. The relativistic RKD follows then from a consistency consideration, i.e.\ from substituting $v^2/\theta^2$ with $\beta(\gamma-1)$. 

Before graphically illustrating these distributions in comparison 
to their non-relativistic counterparts, $N_{\textrm{rSKD}}$ and $N_{\textrm{rRKD}}$ have to be 
computed. Regarding the former, following the calculations in \citet{Xiao-2006}, analytic expressions for the moments of the isotropic rSKD can be found, of which the first few are explicitly given in Appendix~\ref{sec:moments}. The normalization constant is determined by the first moment, i.e.\ the particle four-flow
\begin{equation}
    M_1^\mu=\int \frac{\dd^3p}{p^0}p^\mu f(p), \label{eq:four-flow}
\end{equation}
with $p^0=\sqrt{\mathbf{p}^2+(mc)^2}$ and ${\dd^3p}/{p^0}$ being 
the Lorentz-invariant volume element in phase space, see Appendix~\ref{sec:volume_element} \citep[and also][]{DeGroot, stewart}. According to a usual convention, greek indices denote spacetime 
components of tensors or four-vectors, respectively, and run from 
$0$ to $3$.

From the zeroth component, $M_1^0$, we obtain the number density $n$:
\begin{align}
    n&=\int f_{\textrm{rSKD}}(p)\,\dd^3p =4\pi(mc)^3(\mathrm{I}+\mathrm{II})\,n\,N_{\mathrm{rSKD}} \nonumber
\end{align}
where the second equality is valid for isotropic distributions. 
The auxiliary expressions $\mathrm{I}$ and $\mathrm{II}$ are given by 
\begin{subequations}
\begin{align}
\mathrm{I} &= 2^{1/2}\left(\frac{\kappa}{\beta}\right)^{5/2}B\left(\frac{5}{2},\kappa-2\right) \nonumber \\ 
&\quad\times {}_2F_1\left(-\frac{1}{2},\frac{5}{2};\kappa+\frac{1}{2};1-\frac{\kappa}{2\beta}\right), \quad \kappa>2, \label{rSKD_I}
\end{align}
\begin{align}
\mathrm{II} &= 2^{1/2}\left(\frac{\kappa}{\beta}\right)^{3/2}B\left(\frac{3}{2},\kappa-1\right)\nonumber \\
&\quad\times {}_2F_1\left(-\frac{1}{2},\frac{3}{2};\kappa+\frac{1}{2};1-\frac{\kappa}{2\beta}\right), \quad \kappa>1, \label{rSKD_II}
\end{align}
\end{subequations}
which correspond to $4\pi I_S(0,0)$, see Appendix \ref{sec:moments}.
The sum $\mathrm{I}+\mathrm{II}$ can be further simplified by rewriting the Beta functions in terms of Gamma functions,
\begin{align*}
  \mathrm{I}+\mathrm{II} =&
         \sqrt{2}\left(\frac{\kappa}{\beta}\right)^{3/2}\frac{\sqrt{\pi}}{2}
         \frac{\Gamma\left(\kappa-2\right)}{\Gamma\left(\kappa+\frac{1}{2}\right)}\\
       &
         \left\{\frac{3\kappa}{2\beta} {}_2F_1\left(-\frac{1}{2},\frac{5}{2};\kappa+\frac{1}{2};1-\frac{\kappa}{2\beta}\right)\right.\\
     &\left.    +(\kappa-2) {}_2F_1\left(-\frac{1}{2},\frac{3}{2};\kappa+\frac{1}{2};1-\frac{\kappa}{2\beta}\right)\right\},
\end{align*}
and upon making use of Eq.\,(15.2.19) in \citet{abramowitz},
\begin{align*}
  &(b-a)(1-z){}_{2}F_{1}(a,b;c;z)-(c-a){}_{2}F_{1}(a-1,b;c;z)\\
  &+(c-b){}_{2}F_{1}(a,b-1;c;z)  = 0,
\end{align*}
one finds
\begin{align*}
    \mathrm{I}+\mathrm{II} =&
         \sqrt{\frac{\pi}{2}}\left(\frac{\kappa}{\beta}\right)^{3/2}
          \frac{(\kappa+1)\Gamma\left(\kappa-2\right)}{\Gamma\left(\kappa+\frac{1}{2}\right)}\\
          &\times{}_2F_1\left(-\frac{3}{2},\frac{5}{2};\kappa+\frac{1}{2};1-\frac{\kappa}{2\beta}\right).\\
 \end{align*}
Finally, we conclude for the normalization constant
\begin{align}
    N_{\textrm{rSKD}}&=\frac{1}{(2\pi)^{3/2}(mc)^3}\left(\frac{\beta}{\kappa}\right)^{3/2}\frac{\Gamma(\kappa+\frac{1}{2})}{(\kappa+1)\Gamma(\kappa-2)}\nonumber\\
        &\quad\times \frac{1}{{}_2F_1(-\frac{3}{2},\frac{5}{2};\kappa+\frac{1}{2};1-\frac{\kappa}{2\beta})}. \label{eq:N_rSKD}
\end{align}
With the latter we are now in the position to compute the rSKDs and 
to compare them with their non-relativistic counterparts. 
For the sake of better comparability, we introduce scaled momentum, $\tilde{p}=p/(mc)$, and velocity or speed variables, $\tilde{v}=v/c$, which retain the structure of the distributions, and plot the functions in terms of those.

\begin{figure}
\begin{minipage}{0.504\textwidth}
\includegraphics[width=\textwidth]{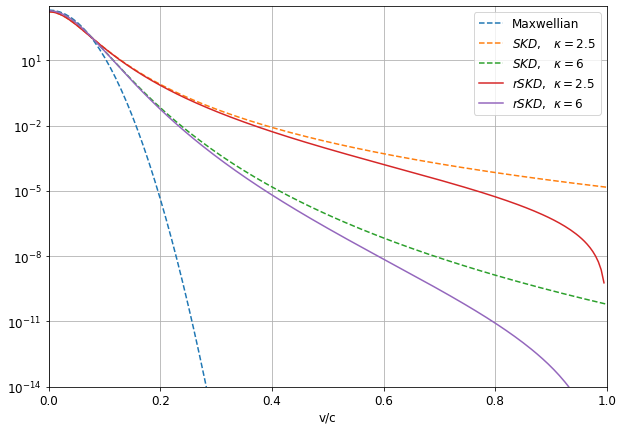}
\end{minipage}
\begin{minipage}{0.496\textwidth}
\includegraphics[width=\textwidth]{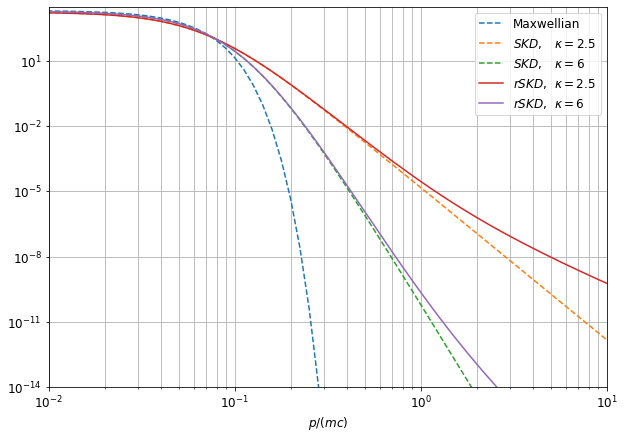}
\end{minipage}
\caption{Non-relativistic SKDs (dashed lines) and relativistic rSKDs (solid lines) are plotted versus speed $v$ scaled by $c$ (upper panel) and versus momentum $p$ scaled by $mc$ (lower panel) for $\beta=10^3$ corresponding to $\theta\approx0.04\,c$, i.e.\ in the non-relativistic regime.}
\label{fig:vgl_skd_rskd}
\end{figure}

In Figure \ref{fig:vgl_skd_rskd} we compare two rSKDs with 
$\beta = 10^3$ corresponding to $\theta \approx 0.04\,c$, i.e.\ in 
the non-relativistic regime, with their non-relativistic counterparts 
for $\kappa = 2.5$ and $\kappa = 6$. 
Both distributions show the same behavior at low scaled speeds $\tilde{v}$ or momenta $\tilde{p}$, respectively, i.e.\ that of the corresponding Maxwellian. With increasing values, a likewise increasing difference arises until the rSKDs fall off with a relativistic cut-off in velocity space. In momentum space, the non-relativistic distributions decrease faster than their relativistic counterparts since the same value for $\tilde{p}$ corresponds to different values of $\tilde{v}$ depending on which framework -- relativistic ($p=\gamma m v$) or non-relativistic ($p=mv$) -- is used.

Interestingly, from the above normalization constant $N_{\textrm{rSKD}}$, 
particularly its leading-order term (\ref{rSKD_I}), follows the constraint $\kappa>2$ for the integral to converge \citep[see also][]{Xiao-2006}, which is an even stronger limitation than in the non-relativistic case, where $\kappa>3/2$ is required for the second-order moment to exist. An alternative and perhaps more evident way to reveal this constraint is to consider the limit $p\rightarrow\infty$, in which the integrand takes the form $p^2\left[1+\frac{\beta}{\kappa}(\gamma-1)\right]^{-\kappa-1}\approx p^2/p^{\kappa+1}$, so that the integral fails to converge for $\kappa\leq 2$. This limitation yet again emphasizes the need for a distribution that is physically consistent, since on the one hand for higher moments of the rSKD the lower boundary for valid $\kappa$ values even increases and since on the other hand power laws with $\kappa \leq 2$ are actually observed. 

The relativistic regularized $\kappa$-distribution (rRKD) offers a  resolution of this problem for the exponential term ensures a cut-off such that the moments remain finite for all $\kappa>0$. Since in the case of the rRKDs we could not find an analytic expression for the normalization constant $N_{\mathrm{rRKD}}$ it was computed numerically.

\begin{figure}
    \centering
\includegraphics[width=0.5\textwidth]{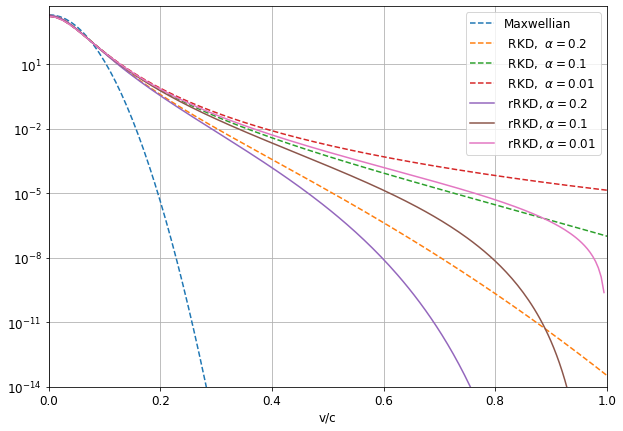}
\includegraphics[width=0.5\textwidth]{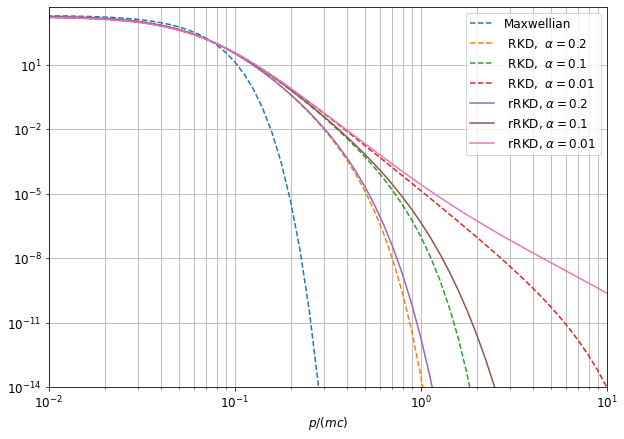}
    \caption{As Fig.~\ref{fig:vgl_skd_rskd} but for the RKDS and rRKDs. For the sake of readability only values for $\kappa=2.5$ are displayed for different cut-off parameters $\alpha$. Again, $\beta=10^3$, $\theta\approx 0.04\,c$, respectively.}
    \label{fig:vgl_rkd_rrkd}
\end{figure}

Using the same representations as in Figure~\ref{fig:vgl_skd_rskd}, 
Figure~\ref{fig:vgl_rkd_rrkd} illustrates two RKDs and the 
corresponding rRKDs. Note that for $\alpha=0.01$ a RKD is displayed (dotted red curve) which is strictly speaking ill-defined because the restriction $\alpha>\theta/c$ must hold for a proper cut-off to occur \citep[see][]{Scherer-etal-2019a}. However, we demonstrate this way that in the case of low $\alpha$ values -- and therefore cut-offs -- the rRKD is an appropriate choice since even though a pronounced tail is formed, the particle speeds are confined to $v<c$. This can potentially be useful for describing power law slopes due to the acceleration of particles.

\begin{figure}
    \centering
    \includegraphics[width=0.5\textwidth]{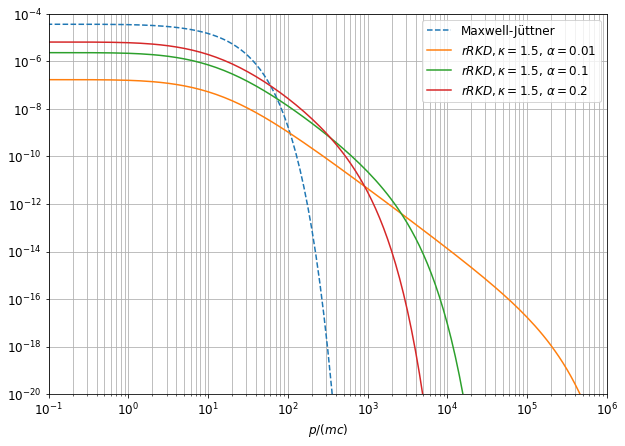}
    \caption{Three rRKDs for $\beta=0.1$ and $\kappa=1.5$. The lower the value of $\alpha$, the later the exponential cut-off occurs.}
    \label{fig:rRKD}
\end{figure}

In order to demonstrate the extended applicability of rRKDs as compared to rSKDs, Figure~\ref{fig:rRKD} displays the former for three different cut-off parameters $\alpha$ and $\kappa=3/2$, a value for which the rSKD cannot be defined.

Briefly turning back to the particle four-flow, we note that, as $M_1^\mu$ and $p^\mu$ both transform as four-vectors while ${\dd^3p}/{p^0}$ is a Lorentz-invariant scalar, the distribution function as well must be a scalar and thus invariant under a Lorentz transformation. For a more thorough evidence of the co-variant 
character of the distribution function, we refer to textbooks such as \citet{DeGroot}.

\section{Limits of the relativistic $\kappa$-distributions}
\subsection{Non-relativistic non-equilibrium distributions}
The functions fulfil the requested conditions at low speeds
\begin{eqnarray}
\beta(\gamma-1)\approx\frac{mc^2}{kT}\left(1+\frac{v^2}{2c^2}-1\right)=\frac{2v^2}{mkT}=\frac{v^2}{\theta^2}
\end{eqnarray}
%
%and high $\kappa$ values, for the expression within the outer brackets in Eq.\,(\ref{eq:rSKD}) just restores the exponential function. 
because, by means of the scaled momentum, we can rewrite the ratio $v^2/\theta^2=\beta\tilde{p}^2/2$ within Eq.\,(\ref{eg:MBD})-(\ref{eq:RKD}) and by using the Taylor expansion $(\gamma-1)=\sqrt{1+\tilde{p}^2}-1\approx\tilde{p}^2/2$ the recovery of the non-relativistic distributions becomes apparent immediately. Here we used the relation $\beta=2c^2/\theta^2$, which just follows from the definitions and from which we may see that $\theta$ can only be an appropriate parameter in the non-relativistic framework. 

Note that $N_{\textrm{rSKD}}$ in (\ref{eq:N_rSKD}) also reduces in 
the non-relativistic limit $\beta \gg 1$, upon using (15.1.20) in 
\citet{abramowitz}, to the normalization factor given in 
\citet{Fahr-Heyl-2020} for that case.

\subsection{Relativistic equilibrium distributions} 
\begin{figure}
    \centering
    \includegraphics[width=0.5\textwidth]{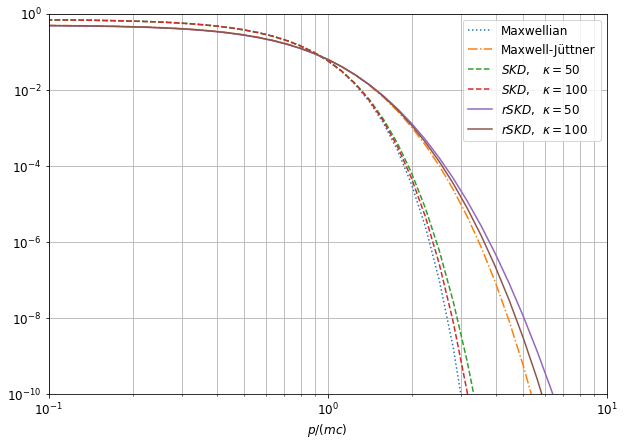}
    \caption{Comparison of the SKD and the rSKD in the limit of high $\kappa$ values. As can be seen, the latter approaches the Maxwell-Jüttner distribution function well. The plot shows the case for $\beta=5$ ($\theta=0.6$\,c).}
    \label{fig:great_kappas}
\end{figure}

With the normalization constant (\ref{eq:N_rSKD}) the rSKD correctly reduces to a MJD in the limit $\kappa\rightarrow\infty$ as is shown in the following. Starting with the integral representation of the hypergeometric function \citep[see (15.3.1) in][]{abramowitz},
\begin{align*}
    {}_2F_1(a,b;c;z)&= \frac{\Gamma(c)}{\Gamma(b)\Gamma(c-b)} \\ &\quad\times\int_0^1  t^{b-1}(1-t)^{c-b-1}(1-tz)^{-a} \,\dd t, 
\end{align*}
and changing the integral limits with the help of
$\int_\alpha^\beta f(x)\dd x=(\beta-\alpha)\int^\infty_0f\left(\frac{\alpha+\beta t)}{1+t}\right)\frac{\dd t}{(1+t)^2}$ \citep[see (3.021) in][]{gradshteyn}, the above expression can be converted to
\begin{align}
    {}_2F_1(a,b;c;z) &= \frac{\Gamma(c)}{\Gamma(b)\Gamma(c-b)} \nonumber \\ 
        &\quad\times \int_0^\infty t^{b-1}(1+t)^{a-c}(1+t[1-z])^{-a}\dd t. \label{eq:2F1_alternative}
\end{align}
This can be applied to the hypergeometric function within the constant $N_{\textrm{rSKD}}$, which we first rewrite according to Eq.\,(15.3.5) in \citet{abramowitz}, 
\begin{align*}
    {}_2F_1&\left(-\frac{3}{2},\frac{5}{2};\kappa+\frac{1}{2};1-\frac{\kappa}{2\beta}\right)\\
    & \qquad= \left(\frac{\kappa}{2\beta}\right)^{-5/2}{}_2F_1\left(\kappa+2,\frac{5}{2};\kappa+\frac{1}{2};1-\frac{2\beta}{\kappa}\right),
\end{align*}
so that we end up with
\begin{align*}
    N_{\textrm{rSKD}}&=\frac{2}{\pi^{3/2}(mc)^3}\left(\frac{\beta}{\kappa}\right)^4\frac{\Gamma(\frac{5}{2})}{(\kappa+1)}  \\
    &\quad\times\left(\int_0^\infty\dd t\,t^{3/2}(1+t)^{3/2}\left(1+t\frac{2\beta}{\kappa}\right)^{-\kappa-2}\right)^{-1}.
\end{align*}
By performing the limit $\kappa\rightarrow\infty$, one obtains the exponential function
$\lim_{\kappa\rightarrow\infty}\left(1+t\frac{2\beta}{\kappa}\right)^{-\kappa-2}=\exp\left(-2t\beta\right)$ and we can express the resulting integral in terms of a modified Bessel function
\begin{align*}
    K_\nu(xz)&=\sqrt{\frac{\pi}{2z}}\frac{x^\nu e^{-xz}}{\Gamma(\nu+\frac{1}{2})}\int^\infty_0 e^{-xt}t^{\nu-\frac{1}{2}}\left(1+\frac{t}{2z}\right)^{\nu-\frac{1}{2}} \dd t
\end{align*}
\citep[see (8.432) in][]{gradshteyn}, with which we conclude
\begin{align*}
    \lim_{\kappa\rightarrow\infty}N_{\textrm{rSKD}}&=\frac{\beta e^{-\beta}}{4\pi(mc)^3K_2(\beta)}.
\end{align*}
The limit of the complete distribution function, accordingly, reads
\begin{align*}
    \lim_{\kappa\rightarrow\infty}f_{\textrm{rSKD}}=&n\frac{\beta e^{-\beta}}{4\pi(mc)^3K_2(\beta)} \lim_{\kappa\rightarrow\infty}\left[1+\frac{\beta (\gamma-1)}{\kappa}\right]^{-\kappa-1}  \\
     =&n\frac{\beta e^{-\beta}}{4\pi(mc)^3K_2(\beta)}e^\beta\exp(-\beta\gamma) \\ =& f_{\textrm{MJD}}. 
\end{align*}
Figure~\ref{fig:great_kappas} illustrates that indeed the rSKD converges to the Maxwell-Jüttner distribution in the limit of high $\kappa$ values. 

Just as its non-relativistic counterpart, the rRKD reduces to the corresponding rSKD in the limit $\alpha\rightarrow 0$, so that in this case the above discussed properties also hold for the former distribution function. Generally, in the limit $\kappa\rightarrow\infty$, the rRKD corresponds to a MJD of the form $\propto \exp\left(-\beta\gamma\left(1+\alpha^2\right)\right)$.
\section{Pressures}
With the relativistic distributions at hand, we may now consider their  moments. Besides the number density considered in section~III, an 
often used macroscopic quantity is the scalar pressure $P$. It is defined as one third of the trace of the pressure tensor \citep[e.g.,][]{balescu}. Expressing this in terms of momentum-space variables, in the non-relativistic case yields
\begin{equation}
P_{nr}=\frac{1}{3m}\int  p^2f(p)\,\dd^3p.
\end{equation}
Therefore, the corresponding pressures to the above non-relativistic isotropic distribution functions are
\begin{subequations}
\begin{align}
    P_{\mathrm{MBD}}&= \frac{1}{2}nm\theta^2 \label{eq:P_MB},\\
    P_{\mathrm{SKD}}&= \frac{1}{2}nm\theta^2\frac{\kappa}{\kappa-\frac{3}{2}} \label{eq:P_SKD},\\
    P_{\mathrm{RKD}}&= \frac{1}{2}nm\theta^2\kappa\frac{U(\frac{5}{2},\frac{5}{2}-\kappa,\alpha^2\kappa)}{U(\frac{3}{2},\frac{3}{2}-\kappa,\alpha^2\kappa)}.
\end{align}
\end{subequations}

In the framework of special relativity, the pressure in the local rest frame is given by the spatial, i.e.\ non-zero, diagonal elements of the energy momentum tensor 
\begin{equation}
    T^{\mu\nu}=c\int \frac{d^3p}{p^0}p^\mu p^\nu f(p).
\end{equation}
 
If we again assume the simplest case of isotropic plasmas, this leads us to the scalar pressure
\begin{equation}
    P_{r}=\frac{1}{3}c\int\frac{\dd^3p}{p_0}p^2f(p). \label{eq:P_rel}
\end{equation}

Inserting the MJD, substituting $p/(mc)=\sinh x$ and using Eq.\,(9.6.23) in \citet{abramowitz}, yields the associated pressure
\begin{equation}
    P_{\mathrm{MJD}}=\frac{1}{\beta}nmc^2,
\end{equation}
which equals $P_{\mathrm{MBD}}$ if we recall that $\beta=2c^2/\theta^2$. Since the ideal gas law $P=n k_B T$ holds in both the non-relativistic and the relativistic case, this consistency is just what we expect and what serves as a check for our formulas.

According to Appendix \ref{sec:moments}, an analytic expression for the case of the isotropic rSKD can as well be found by taking the trace of the second moment $M_2$ and multiplying with the pre-factor as indicated in Eq.\,(\ref{eq:P_rel}), 
\begin{align}
    P_{\textrm{rSKD}}&=\frac{1}{\beta}nmc^2\frac{\kappa(\kappa-\frac{1}{2})}{(\kappa+1)(\kappa-3)}\nonumber\\
    &\quad\times\frac{{}_2F_1(-\frac{3}{2},\frac{5}{2};\kappa-\frac{1}{2};1-\frac{\kappa}{2\beta})}{{}_2F_1(-\frac{3}{2},\frac{5}{2};\kappa+\frac{1}{2};1-\frac{\kappa}{2\beta})},\quad \kappa>3. 
\end{align}

Because for the calculation of higher-order moments additional factors of $p$ appear, the restriction of $\kappa$ extends to even higher values. Here, in the case of the second moment, $\kappa > 3$ is required for the integral to converge, whereas the restriction in the purely non-relativistic case was only $\kappa> 3/2$. 

In the non-relativistic limit, $\beta\gg1$, $P_{\mathrm{rSKD}}\approx \frac{1}{\beta}nmc^2\frac{\kappa}{\kappa-\frac{3}{2}}$, where we used Eq.\,(15.1.20) in \citet{abramowitz}, which is indeed the same value as in Eq.\,(\ref{eq:P_SKD}). Applying Eq.\,(15.3.4) in \citep{abramowitz} and then taking the limit $\kappa\rightarrow\infty$, again allows us to use Eq.\,(15.1.20) and we obtain $\lim_{\kappa\rightarrow\infty}P_{\mathrm{rSKD}}=\frac{1}{\beta}nmc^2=P_{\mathrm{MJD}}$.

\begin{figure}
    \centering
    \includegraphics[width=0.5\textwidth]{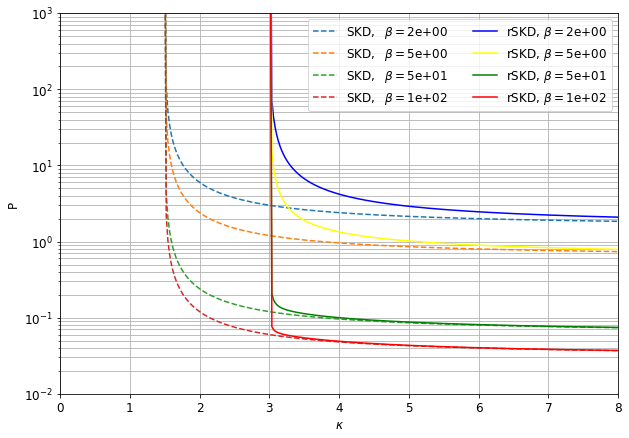}
    \caption{Comparison of pressures calculated in a purely non-relativistic (dashed lines) and a relativistic (solid lines) manner of SKD-distributed plasmas.}
    \label{fig:pres_skd+rskd}
\end{figure}

In the corresponding plots, we display the scaled pressures $\tilde{P}=3P/(m^4c^5)$ vs.\ $\kappa$. Figure \ref{fig:pres_skd+rskd} compares the pressures obtained either by use of the fully non-relativistic theory or the relativistic formulation.
The restriction of the rSKD to $\kappa>3$ emerges as the pressure diverges for lower values of $\kappa$. 

\begin{figure}
    \centering
    \includegraphics[width=0.5\textwidth]{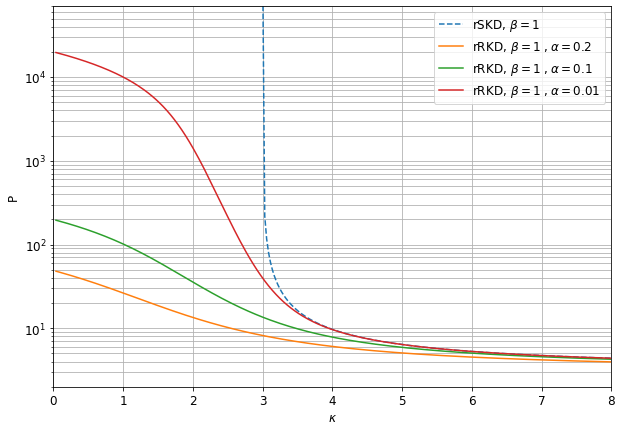}
    \caption{Pressures of rRKDs for $\beta=1$ and three different cut-off parameters $\alpha$. The smaller $\alpha$, the more the function approaches the divergence that the corresponding rSKD pressure performs.}
    \label{fig:pres_rSKD+rRKD}
\end{figure}

\begin{figure}
    \begin{minipage}{0.5\textwidth}
    \includegraphics[width=\textwidth]{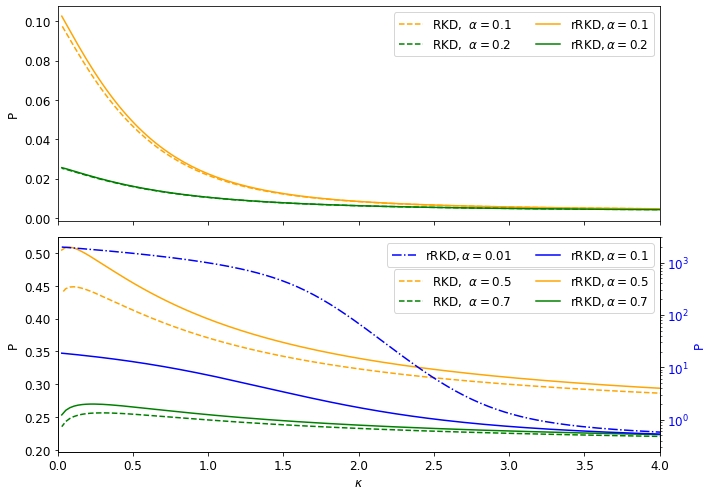}
    \end{minipage}
    \caption{The pressures of RKDs and rRKDs are compared for different cut-off parameters $\alpha$. In the upper panel, our parameter is taken to be $\beta=10^3$ ($\theta=0.04$\,c), in the lower panel, $\beta=10$ ($\theta=0.45$\,c). Note that the vertical axis on the right belongs to the blue-coloured rRKDs.}
    \label{fig:pres_rkd+rRKD}
\end{figure}

Turning to the rRKDs, figure \ref{fig:pres_rSKD+rRKD} indicates that such limitations can be removed for sufficient high parameters $\alpha$.
Figure \ref{fig:pres_rkd+rRKD} compares the pressures of regularized-$\kappa$ distributed plasmas in the purely non-relativistic to the relativistic description.
Within the scope of $\alpha$-values, for which the RKD is defined, i.e.\ $\alpha>\theta/c$ \citep[see][]{Scherer-etal-2019a}, the pressures obtained in the non-relativistic framework may be slightly underestimated, viz.\ the more, the lower $\beta$ and the lower $\alpha$. Since the deviations between both descriptions seem to be small, the use of the non-relativistic RKD might, nevertheless, be well justified. As soon as the cut-off parameter $\alpha$ decreases below the value $\theta/c$, the non-relativistic RKD is not valid anymore and the rRKD could be the preferred choice, since, as indicated by figure \ref{fig:vgl_rkd_rrkd}, the latter remains well defined. The lower panel of figure \ref{fig:pres_rkd+rRKD} displays two such distributions, for which no non-relativistic correspondence exists.
%(Due to the apparent deviations, the plots allude to the fact that the pressures obtained in the non-relativistic framework may be underestimated, viz. the more, the higher $\beta$ and the lower $\alpha$.)

\section{Summary}
%Several attempts have been made to formulate relativistic $\kappa$-distribution functions, e.g. \textcite{Xiao-2006}, \textcite{Treumann-Baumjohann-2018}, \textcite{Fahr-Heyl-2020}. In the latter, the authors come to a different result for the second moment of their distribution function as they replace the energy expression within $\int E_{kin}f(E_{kin})\, \dd^3v$ but choose to integrate over velocity space for they predominantly study the sub-relativistic regime. \textcite{Treumann-Baumjohann-2018}, on the other hand, introduce a Lorentzian distribution function that is valid for ultra-relativistic energies only and, accordingly, express the particle energy as $E\approx pc$ since they apply it to high-energy cosmic ray spectra.
%The former author uses the same structure for the isotropic relativistic standard $\kappa$ distribution, though their weighting parameter $\theta$ differs from ours.  \\
%According to \citet{Lazar-2016}, our rather heuristic choice of $\beta$ corresponds to the "Kappa-B" interpretation since this parameter determines the thermal speed $\theta$ which then is equal for the $\kappa$-distributions and their associated Maxwellian. This, consequently, leads to a $\kappa$-dependent temperature. 

In this paper, we derived the special relativistic generalization of isotropic kappa distributions, based on the requirement that in the non-relativistic limit the familiar distributions are recovered. We find that extending the Olbertian or standard kappa distributions into the framework of a relativistic description leads to a stronger restriction of $\kappa$-values than in the non-relativistic case. This underlines the need to formulate distributions that do not result in divergences of velocity moments and are valid for all positive kappa values. 

The relativistic regularized kappa distributions offer, just as their non-relativistic counterparts, physical consistency for they completely remove such constraints. These newly derived co-variant distributions only allow for speeds $v<c$ and are, thereby, removing also all remaining limitations regarding the applicability of non-relativistic regularized kappa distributions. In this sense, the relativistic regularized kappa distributions complete the theory of this family of distribution functions, and pave the way to a consistent treatment of highly energetic particle populations characterized by power-laws in velocity or momentum as frequently occuring in various space- and astrophysical systems.

\begin{appendix}
\section{Lorentz-invariant volume element}\label{sec:volume_element}
A coordinate-independent volume element $\dd\Omega_P$ in phase space is given by
\begin{align}
\dd\Omega_P=\sqrt{|g|}\,\dd^4 p &= \sqrt{|g|}\,\dd p^0\wedge\dd p^1\wedge\dd p^2\wedge\dd p^3 \nonumber \\
&=\sqrt{|g|}\,\epsilon_{abcd}\dd p^{a}\otimes\dd p^{b}\otimes\dd p^{c}\otimes\dd p^{d} \nonumber \\
&= \sqrt{|g|}\,\dd p^{0}\otimes\dd p^{1}\otimes\dd p^{2}\otimes\dd p^{3}\label{eq:dV_p}
\end{align}
\citep[cf.][]{stewart, carroll}, where $\wedge$ denotes the wedge product, which is the antisymmetrized tensor product of two differential forms, and $\epsilon_{abcd}$ the Levi-Civita symbol
\begin{equation}
\epsilon_{abcd}=
\begin{cases} +1, \qquad &\mathrm{if} \,abcd \,\mathrm{is\, an\, even\, permutation\, of\,} 0123 \\ 
-1, &\mathrm{if}\, abcd \,\mathrm{is\, an\, odd\, permutation\, of\,} 0123\\ 
0, &\mathrm{else}.\end{cases}  
\end{equation}
Since we do not deal with curved space-time and confine ourselves to special relativity, $g$ is the determinant of the Minkowski metric $g_{\mu\nu}=\textrm{diag}(1,-1,-1,-1)$ and $\sqrt{|g|}=1$. Greek indices run from 0 to 3. (\ref{eq:dV_p}) can also be expressed in terms of the ordinary three-dimensional momentum space element by restricting to the mass-shell
\begin{equation}
    p^\mu p_\mu = (p^0)^2-\bm{p}^2=m^2c^2 \label{eq:mass_shell}
\end{equation}
which is, roughly speaking, the physically accessible part of momentum space for particles with rest mass $m$. Therefore, one can define
\begin{equation}
\dd V_p\equiv 2\textrm{H}(p^0)\delta(p_\mu p^\mu-m^2c^2)\,\dd^4p \label{eq:vol_element}
\end{equation} 
in which $\textrm{H}(p^0)=\left\{\begin{array}{ll} 1, & p^0>0 \\ 0, & \textrm{otherwise}\\ \end{array}\right.$ is the Heaviside step function. \\
With the identity
\begin{equation}
\mathrm{H}(p^0)\,\delta(p_\mu p^\mu-m^2c^2) = \frac{1}{2p^0}\delta\left(p^0-\sqrt{\bm{p}^2+m^2c^2}\right), \nonumber
\end{equation} 
the volume element (\ref{eq:vol_element}) reduces to
\begin{align}
\dd V_p &= \frac{1}{p^0}\delta\left(p^0-\sqrt{\bm{p}^2+m^2c^2}\right)\dd p^0\otimes\dd p^1\otimes\dd p^2 \otimes\dd p^3 \nonumber\\
&=\frac{1}{p^0} \dd^3p,
\end{align}
where in the last step we used that in locally orthonormal coordinates $\dd p^1\otimes\dd p^2\otimes \dd p^3=\dd^3 p$.
The mass shell condition is implicitly carried within the distribution function after evaluation of $\int \delta\left(p^0-\sqrt{\bm{p}^2+m^2c^2}\right) f \dd p^0$.
%The mass shell condition is implicitly carried within the distribution function after evaluation of $\int \delta\left(p^0-\sqrt{\mathbf{p}^2+m^2c^2}\right) f \dd p^0$.

%
\section{Moments of the isotropic rSKD}\label{sec:moments}
When calculating the moments of the isotropic rSKD, integrals of the form $\int (p^0)^h|\mathbf{p}|^j f_{\mathrm{rSKD}}\,\dd^3p$ appear.
From (\ref{eq:mass_shell}), we obtain the relation \begin{equation}
    p^0=\sqrt{\bm{p}^2-m^2c^2} \label{eq:p0}
\end{equation}
and with the substitution $\tilde{E}=\sqrt{1+p^2/(mc)^2}-1$, which actually is the kinetic energy scaled by $mc^2$, $\dd p=(mc)^2\frac{\tilde{E}+1}{\sqrt{\tilde{E}(\tilde{E}+2)}}\dd \tilde{E}$, we can write
\begin{align}
   \int & (p^0)^h|\bm{p}|^j f_{\mathrm{rSKD}}\,\dd^3p \propto (mc)^{h+j+3}nN_{\mathrm{rSKD}} \nonumber\\
    &\times\int (\tilde{E}+1)^{h+1}\left[\tilde{E}(\tilde{E}+2)\right]^{\frac{j+1}{2}}\left[1+\frac{\beta}{\kappa}\tilde{E}\right]^{-(\kappa+1)}\dd \tilde{E} \nonumber\\
   =&(mc)^{h+j+3}nN_{\mathrm{rSKD}}\left(\frac{\kappa}{\beta}\right)^{\kappa+1} \nonumber\\
    &\times\int \sum_{i=0}^{h+1}\binom{h+1}{i}\tilde{E}^{i+\frac{j+1}{2}}(\tilde{E}+2)^{\frac{j+1}{2}}\left[\frac{\kappa}{\beta}+\tilde{E}\right]^{-(\kappa+1)} \dd \tilde{E}\nonumber\\
    =& (mc)^{h+j+3}nN_{\mathrm{rSKD}}  \sum_{i=0}^{h+1}\frac{(h+1)!}{(1+h-i)!i!} 2^{\frac{j+1}{2}}\nonumber \\
    &\times \left(\frac{\kappa}{\beta}\right)^{i+\frac{j+3}{2}}B\left(i+\frac{j+3}{2},\kappa-j-i-1\right) \nonumber\\
    &\times {}_2F_1\left(-\frac{j+1}{2}, i+\frac{j+3}{2};\kappa+\frac{1-j}{2};1-\frac{\kappa}{2\beta}\right) \nonumber \\
    \equiv &\, I_S(h, j),
\end{align}
where we used (3.197) in \citet{gradshteyn}. The restriction $\kappa>h+j$ holds. \\
We explicitely give the first five moments of the isotropic rSKD,
\begin{align*}
    &M_0 = \int \frac{\dd^3p}{p_0}f_{\mathrm{rSKD}}= 4\pi I_S(h=-1, j=0) \\
    &M^\mu_1 = \int \frac{\dd^3p}{p_0}p^\mu f_{\mathrm{rSKD}}=\begin{cases}4\pi I_S(0,0),\quad &\mu=0 \\0, &\mu=1,2,3\end{cases} \\
    &M^{\mu\nu}_2= \int \frac{\dd^3p}{p_0}p^\mu p^\nu f_{\mathrm{rSKD}} =\begin{cases}4\pi I_S(1,0),\quad &\mu=\nu=0 \\ \frac{4\pi}{3}I_S(-1,2), &\mu=\nu \neq 0 \\ 0, &\mathrm{else}\end{cases} \\
    &M^{\mu\nu\rho}_3= \int \frac{\dd^3p}{p_0}p^\mu p^\nu p^\rho f_{\mathrm{rSKD}} \\
        &\;=\begin{cases} 4\pi I_S(2,0), \quad & \mu=\nu=\rho=0\\ \frac{4\pi}{3}I_S(0,2),  & \mu=0, \nu=\rho\neq 0 \, \footnotemark \\ 0, & \mathrm{else} \end{cases} \\
    &M^{\mu\nu\rho\tau}_4= \int \frac{\dd^3p}{p_0}p^\mu p^\nu p^\rho p^\tau f_{\mathrm{rSKD}} \\
        &\;= \begin{cases} 4\pi I_S(3,0), \quad & \mu=\nu=\rho=\tau=0\\ \frac{4\pi}{3}I_S(1,2),  & \mu=\nu=0, \rho=\tau\neq 0  \footnotemark[\value{footnote}] \\
        \frac{4\pi}{5} I_S(-1,4), & \mu=\nu=\rho=\tau\neq 0 \\ \frac{4\pi}{15} I_S(-1,4), & \mu=\nu\neq 0, \rho=\tau\neq 0, \mu\neq\rho  \footnotemark[\value{footnote}] \\ 0, &\textrm{else} \end{cases} \\
    &M^{\mu\nu\rho\tau\chi}_5= \int \frac{\dd^3p}{p_0}p^\mu p^\nu p^\rho p^\tau p^\chi f_{\mathrm{rSKD}} \\   
        &\;= \begin{cases}4\pi I_S(4,0), \quad & \mu=\nu=\rho=\tau=\chi=0\\ \frac{4\pi}{3}I_S(2,2), &\mu=\nu=\rho=0, \tau=\chi\neq 0\footnotemark[\value{footnote}] \\
        \frac{4\pi}{5}I_S(0,4), & \mu=0, \nu=\rho=\tau=\chi\neq 0 \footnotemark[\value{footnote}] \\
        \frac{4\pi}{15} I_S(0,4), &\mu=0, \nu=\rho\neq 0, \tau=\chi\neq0, \nu\neq\tau \footnotemark[\value{footnote}] \\
        0, &\textrm{else} \end{cases}
\end{align*}\footnotetext[\value{footnote}]{or a permutation thereof}
where the pre-factors result from the angle integration.
\end{appendix}

%\begin{acknowledgements}
%...\\
%...\\
%\end{acknowledgements}
%merlin.mbs aipauth4-1.bst 2010-07-25 4.21a (PWD, AO, DPC) hacked
%Control: key (0)
%Control: author (9) reversed initials
%Control: editor formatted (0) differently from author
%Control: production of article title (0) allowed
%Control: page (1) range
%Control: year (1) truncated
%Control: production of eprint (0) enabled
%

%\bibliography{literatur}
%
\end{document}